\documentclass{phep}       

\journal{PHEP}
\vol{xx}

\received{xx January 2025}
\published{xx March 2025}

\usepackage{amsmath}
\usepackage{amssymb}
\usepackage[hidelinks]{hyperref}
\usepackage[left]{lineno}

\def\be{\begin{equation}}
\def\ee{\end{equation}}
\def\bea{\begin{eqnarray}}
\def\eea{\end{eqnarray}}

\renewcommand{\thefootnote}{\fnsymbol{footnote}}

\begin{document}

\title{Search for black holes and sphalerons using novel machine learning techniques at CMS}

\author{Tamas Almos Vami\auno{1}\footnotemark\ and Danyi Zhang\auno{1}  for the CMS Collaboration}
\address{$^1$Department of Physics, University of California, Santa Barbara, CA 93106, USA}

\begin{abstract}
A comprehensive search for microscopic black holes and electroweak sphalerons is presented, using proton-proton collision data collected by the CMS detector during 2016–2018, corresponding to an integrated luminosity of $138~\mathrm{~fb}^{-1}$. A novel tool has been developed to identify collider events with distinct kinematic features, based on the phase-space distance between events. Model-independent limits are set on the cross section of new physics signals producing multiple jets and leptons, which are further interpreted as constraints on black hole and sphaleron production. In the context of models with large extra dimensions, semiclassical black holes with masses below 9.0-11.4~TeV are excluded by this search, significantly extending previous sensitivity. Additionally, a dedicated search for electroweak sphaleron transitions has been performed. An upper limit of 0.0025 is set at 95\% confidence level on the fraction of quark–quark interactions with center-of-mass energy above the nominal threshold of 9~TeV that result in sphaleron transitions.
\end{abstract}

\maketitle


\renewcommand{\thefootnote}{\fnsymbol{footnote}}
\footnotetext{Corresponding author: Tamas.Almos.Vami@cern.ch}

\section{Black holes at CMS?}

Usually, discussions about black holes (BHs) concern cosmic scales involving enormous masses and volumes.
However, the criterion for black hole formation is rather simple: the object must be smaller than its Schwarzschild radius. 
The Schwarzschild radius for an object with mass $M$ is given in Eq.~\ref{eq:schwarzschild}, where $G$ is the gravitational constant and $c$ is the speed of light~\cite{Schwarzschild:1916uq}.

\begin{equation}
    R_S = \frac{2GM}{c^2}
    \label{eq:schwarzschild}
\end{equation}

If we calculate the radius for the Earth, $R_S(M=M_{\mathrm{Earth}})$, we find that compressing our planet roughly to the size of an acorn would be enough for it to become a black hole.

The Large Hadron Collider (LHC) at CERN collides proton beams with the center-of-mass energy of 13~TeV. One could ask the question: within what radius would this energy need to be confined to form a black hole? Following the same calculation we find that  $R_S(M=13~\mathrm{TeV}) \approx10^{-50}~\mathrm{m}$. That is an absurdly small scale: for comparison, the Planck lengths, the smallest measurable unit of length in physics, is $l_{\mathrm{Planck}}\approx10^{-35}~\mathrm{m}$. Thus, we can conclude that the LHC cannot produce black holes, assuming no new physics beyond the Standard Model (SM).

However, if our world has more than $3+1$ dimensions~\cite{Antoniadis:1998ig,Arkani-Hamed:1998jmv,Arkani-Hamed:1998sfv,Randall:1999vf,Randall:1999ee}, denoting the extra dimensions with $n$, the Schwarzschild radius calculation changes to as described in Eq.~\ref{eq:schwarzschild_extra}, where $M_{BH}$ is the mass of the black hole, $M_{D}$ is the Planck scale in $n$ dimensions, and $\Omega_{n+2} = (2\pi)^{\frac{n+3}{2}}/\Gamma(\frac{n+3}{2})$.

\begin{equation}
    R'_S = \left[\frac{16\pi}{(n+2)\Omega_{n+2}} \frac{M_{BH}}{M_D^{n+2}}\right]^{\frac{1}{n+1}}
    \label{eq:schwarzschild_extra}
\end{equation}

With this modification, the $R'_S(M=13~\mathrm{TeV}) \approx10^{-20}~\mathrm{m}$. Based on the Heisenberg uncertainty, $\Delta x=13~\mathrm{TeV}/\hbar c \approx10^{-20}~\mathrm{m}$, the extra-dimensional microscopic black holes are within the LHC reach!

\begin{figure}[h!]
    \centering
    \includegraphics[width=0.99\linewidth]{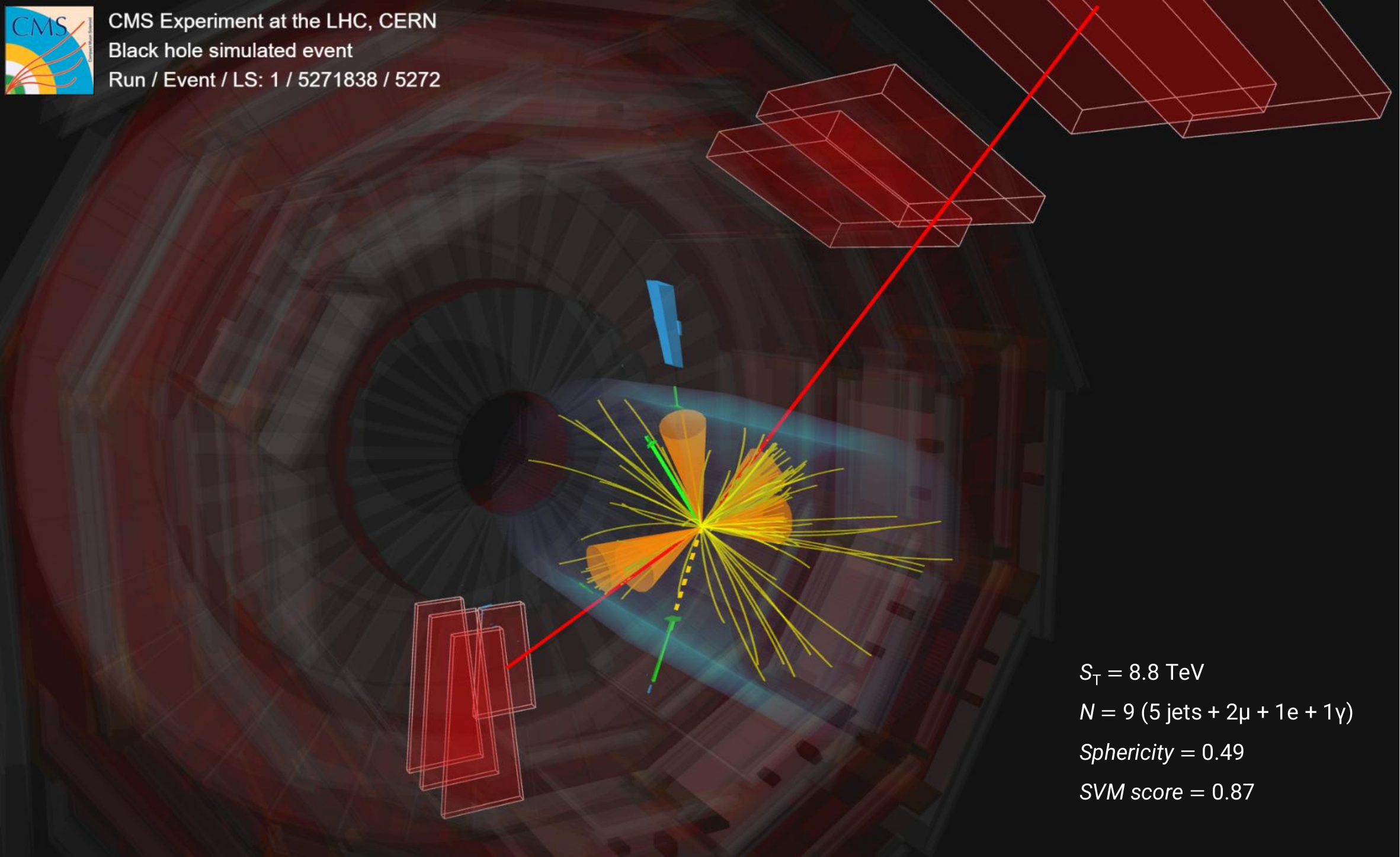}
    \caption{Event display of the final products from a simulated microscopic black hole evaporation at the LHC. The red lines represent muons, the orange cones are jets, the yellow lines are tracks, the green line is an electron, and the blue block is for a photon interaction in the electric calorimeter.}
    \label{fig:event_display_bh}
\end{figure}

However, even if the LHC could produce such objects, they would undergo Hawking radiation. The time $\tau$ it takes to evaporate a black hole is proportional to the cube of its mass, $\tau \propto M^3$. With this the $\tau(M = M_{\mathrm{Earth}}) \approx10^{58}~\mathrm{s}$ and  $\tau(M=13~\mathrm{TeV}) \approx10^{-28}~\mathrm{s}$. So a microscopic black hole at the LHC will immediately decay to high multiplicity, high energy and isotropic objects. The CMS experiment~\cite{CMS:2008xjf} will thus reconstruct these as jets, leptons and missing momentum, as illustrated in Fig.~\ref{fig:event_display_bh}. Isotropic events like these are really hard to separate from SM events, so distinguishing them requires innovative techniques and machine learning tools. The results presented in this work are documented in CMS as \texttt{CMS-PAS-EXO-24-028}~\cite{CMS-PAS-EXO-24-028}.

\section{What are sphalerons?}

\begin{figure}[h!]
    \centering
    \includegraphics[width=0.99\linewidth]{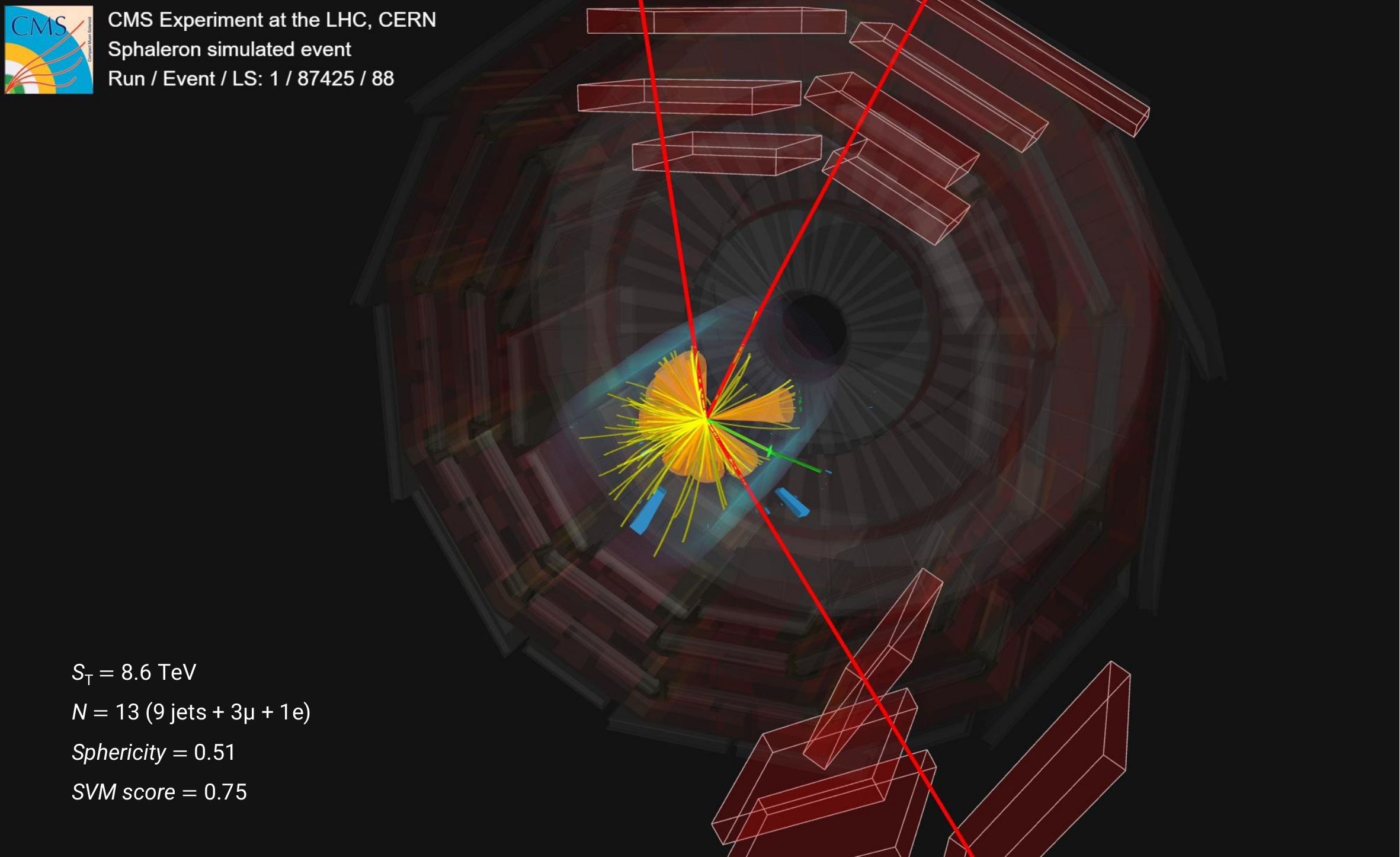}
    \caption{Event display of the final products from a simulated sphaleron process at the LHC.}
    \label{fig:event_display_sph}
\end{figure}

Sphalerons are unstable solutions of the electroweak vacuum within the SM~\cite{tHooft:1976rip,Klinkhamer:1984di}. The cross section of such processes is theorized \cite{Tye:2015tva,Ellis:2016ast,Papaefstathiou:2019djz} to be proportional to a factor called the pre-exponential factor. The periodicity of the vacua is denoted with the Chern-Simons number, $N_{CS}$, while the amplitude is denoted with $E_{\mathrm{sph}}$. A sphaleron transition corresponds to a change in vacuum configuration, i.e., a shift in the $N_{CS}$. We denote the probability of a $\Delta N_{CS} = 1$ change with $p({N_{CS}})$. In these processes, baryon number conservation is violated and a particle from each SM family is created. This mechanism can explain the matter-antimatter asymmetry problem. The final products are similar to the BH evaporation products, displayed in Fig.~\ref{fig:event_display_sph}, thus the same analysis techniques can be applied.

\section{Simulated samples}

We produced black hole signal samples under different assumptions regarding their angular momentum, energy loss, and decay mechanisms. We generated samples labeled with \texttt{B} using the \textsc{BlackMax}~\cite{Dai:2007ki} generator: \texttt{B1} corresponds to non-rotating black holes, \texttt{B2} to rotating ones without graviton emission, and \texttt{B3} to rotating black holes that include energy and momentum loss. We also generated samples labeled with \texttt{C} using the \textsc{Charybdis2}~\cite{Harris:2003db,Frost:2009cf}  generator: \texttt{C1} represents rotating black holes, \texttt{C2} non-rotating ones, \texttt{C3} rotating black holes with the evaporation model described in Refs.~\cite{Creek:2007sy,Creek:2007pw}, \texttt{C4} rotating ones with the Yoshino--Rychkov (YR) model~\cite{Yoshino:2005hi}, which provides a conservative bound incorporating impact parameter effects, \texttt{C5} rotating black holes with a stable remnant model~\cite{Koch:2005ks,Stoecker:2006we,Scardigli:2008jn}, where the remnant is a colorless object akin to a quark--antiquark bound state, and \texttt{C6} non-rotating black holes with a boiling remnant model~\cite{Dimopoulos:2001qe,Gingrich:2008di}, in which the remnant continues to evaporate until the Hawking temperature reaches a maximum value equal to~$M_D$.

We used the \textsc{BaryoGEN}~\cite{Trodden:1998ym} generator to simulate sphaleron processes. We simulated events with $p({N_{CS}})$ values of  0\%, 50\% and 100\%. The meaning of 50\% is that half of the events have $\Delta N_{CS} =-1$ while the other half have $\Delta N_{CS} =1$.

To guide the selection criteria and train machine learning models, we simulated four-flavor Quantum Chromodynamics (QCD) multijet production with \textsc{MadGraph}~\cite{Frederix:2018nkq} at leading order using the MLM matching scheme~\cite{Alwall:2007fs}.

\section{Novel data representation}
In Ref.~\cite{Larkoski:2020thc}, the authors proposed that a collision event can be represented on the phase space manifold $\Pi_N$ of the scattering amplitudes for the $N$ objects in the event. It can be expressed as the product of an $(N-1)$-simplex $\Delta_{N-1}$ and a $(2N-3)$-dimensional hypersphere $S_{2N-3}$, i.e., $\Pi_N \cong \Delta_{N-1} \times S_{2N-3}$. For this study, we set $N=30$ and in case there are fewer objects in the event, we assign zero values to them. Next, explicit global coordinates  of the phase space were introduced, denoted by ${\vec \rho}$ for the simplex and ${\vec v}$ for the hypersphere. Using these coordinates, one can define a distance between two events, $A$ and $B$, denoted as $d_{\Pi}({\vec \rho}_A,{\vec v}_A'; {\vec \rho}_B,{\vec v}_B')$ and expressed in Eq.~\ref{eq:distance_total}, where $d_\Delta$ is the distance on the simplex, $d_S$ is the distance on the hypersphere, $c$ serves as a reweighting factor that gives rise to equally valid phase space distances, $\alpha = \frac{1}{16\pi^2}\left( \frac{c}{4} \right)^{\frac{3-2N}{3N-4}}$ and $\beta = \frac{1}{4\pi^2}\left( \frac{c}{4} \right)^{\frac{N-1}{3N-4}}$. Further details of this metric calculation, we refer to Ref.~\cite{Cai:2024xnt}.

\begin{equation}
    d_{\Pi}({\vec \rho}_A,{\vec v}_A'; {\vec \rho}_B,{\vec v}_B') = 
    \sqrt{ \alpha \, d^2_\Delta({\vec \rho}_A, {\vec \rho}_B) + \beta \, d^2_{S}({\vec v}_A', {\vec v}_B') }
    \label{eq:distance_total}
\end{equation}

\begin{figure}[h!]
    \centering
    \includegraphics[width=1\linewidth]{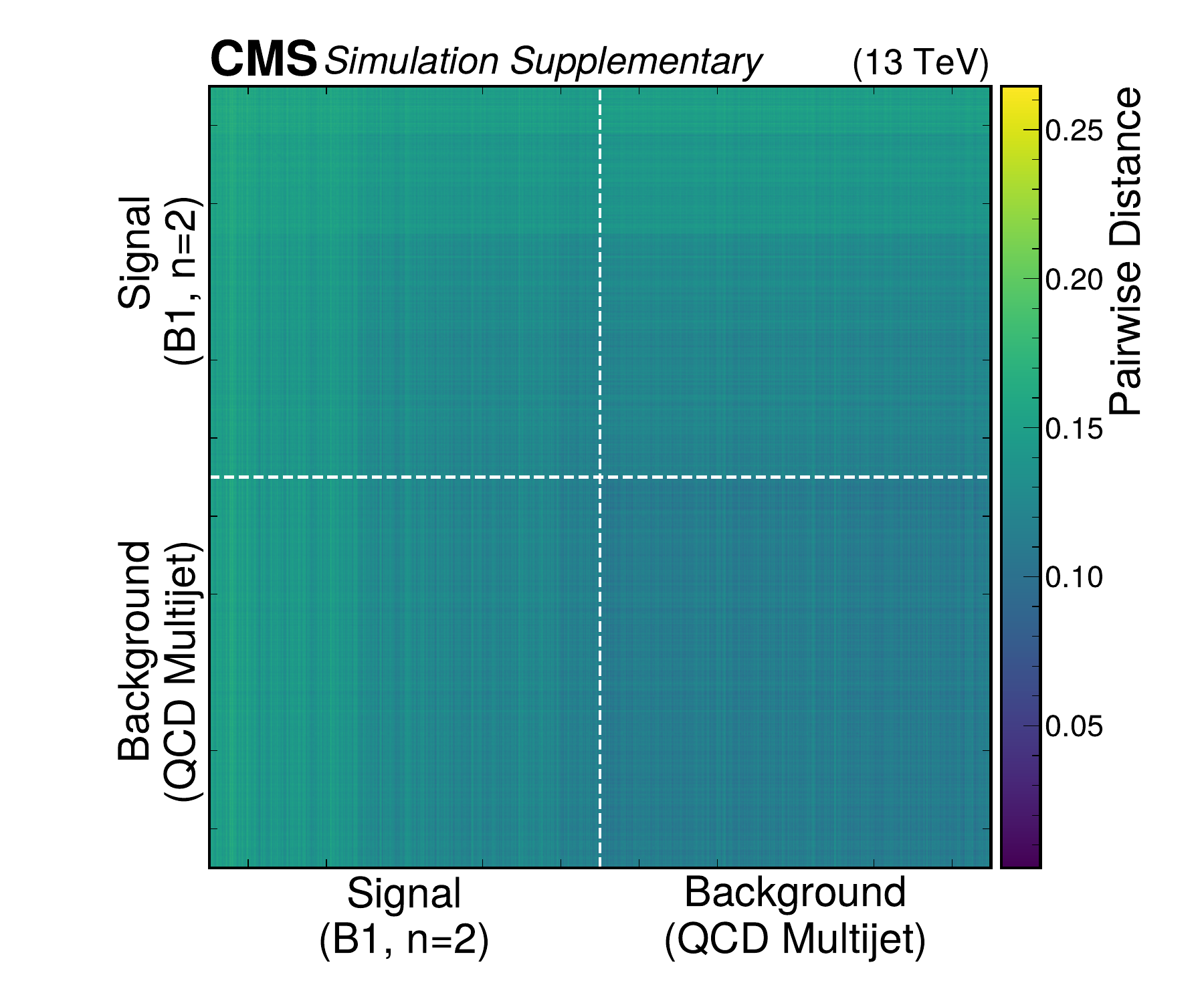}
    \caption{Pairwise distance between 10,000 events in each category of signal (BH with mixture mass points) and background (QCD multijets).}
    \label{fig:pairwise_distance}
\end{figure}

We calculated the pairwise distance between 10,000 events from a mixture of BH mass points with the \texttt{B1} model assuming $n=2$ as signal, and the QCD multijet sample as background. This is displayed as a two-by-two matrix of 10,000 points in each block in Fig.~\ref{fig:pairwise_distance}. While we can see that background-to-background comparisons yield smaller distances, generally it is really hard for humans to discriminate background vs signal based on this figure. 

\section{Machine learning tools}

\begin{figure}[h!]
    \centering
    \includegraphics[width=0.8\linewidth]{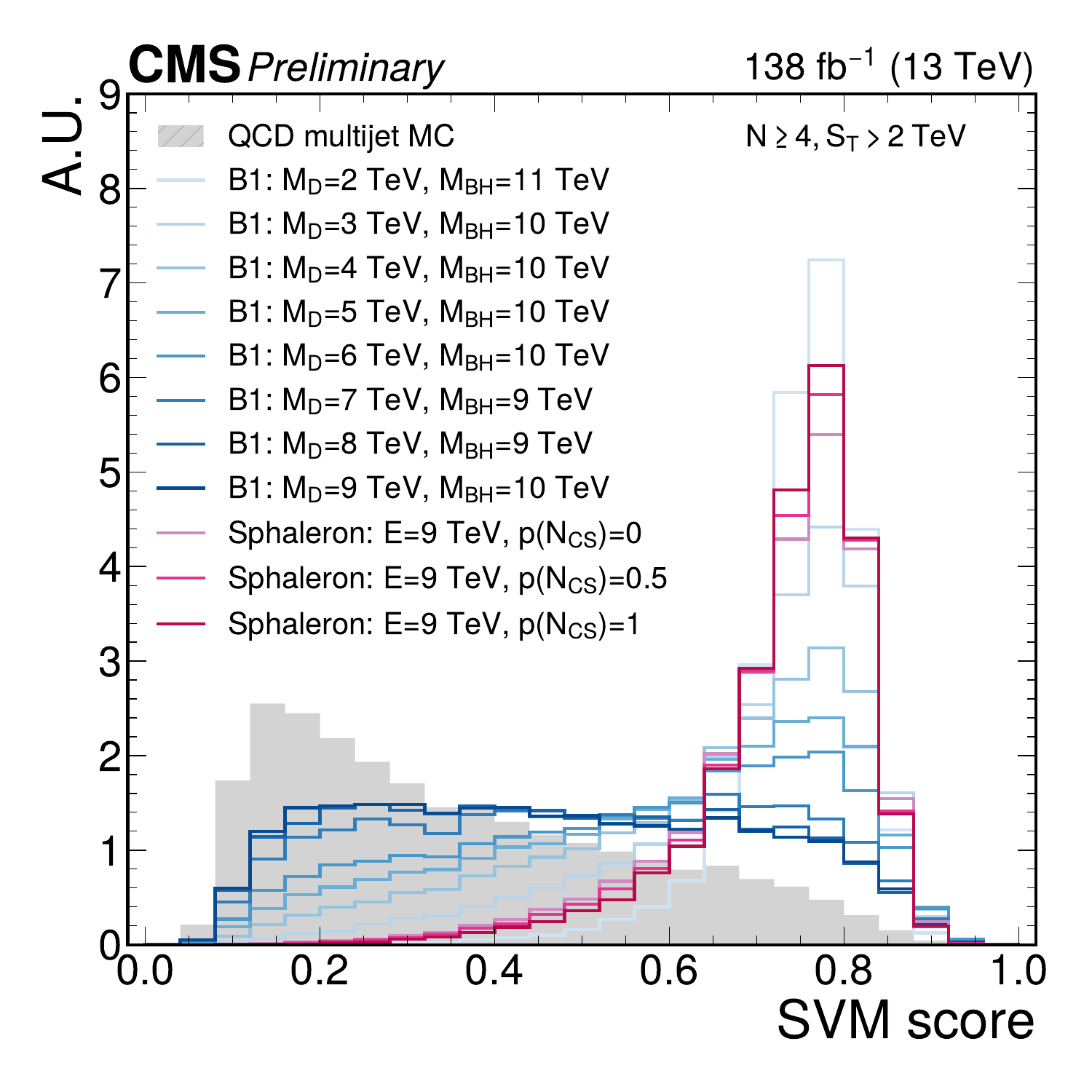}
    \caption{SVM score distributions for simulated QCD multijets and selected black hole (with $n = 2$) and sphaleron models.}
    \label{fig:svm_with_wo_sphericity}
\end{figure}

To overcome the challenges of interpreting high-dimensional data, we employ advanced machine learning techniques. Specifically, we used a Support Vector Machine (SVM)~\cite{Cortes:1995hrp}, which is a distance-based tool that finds the optimal hyperplane to classify signal from background. Since our input representation is inherently geometric, the SVM provides a natural and interpretable choice. The output of the SVM is a probability of how signal-like a given event is. The distribution of the SVM score is displayed in Fig.~\ref{fig:svm_with_wo_sphericity}. We optimized the selection criteria on this variable using the Punzi figure of merit~\cite{Punzi:2003bu}, and found that a single global selection value at 0.63 is sufficient for all mass points, and certainly simplifies the analysis. Varying this threshold between 0.5 and 0.7 changes the signal yield by only 4–25\%, depending on $M_{\text{BH}}$. Unlike other ML approaches, here the input is a physically meaningful distance measure, making the results more interpretable than those from, for example, a neural network.

\section{Further observables}

We require all reconstructed objects to pass standard quality criteria and to have a transverse momentum $p_T$ of at least $70$~GeV. The number of objects after these selections is denoted with $N$ and for the final results we consider only events with $N > 3$.

Since signal events are expected to be highly spherical, we also used sphericity as a selection variable. Sphericity is calculated as shown in Eq.~\ref{eq:sphericity}.
\begin{equation}
    S = \frac{2 \lambda_2}{\lambda_2 + \lambda_1},
    \label{eq:sphericity}
\end{equation}
where $\lambda_1$ and $\lambda_2$ are the eigenvalues of the transverse momentum tensor $S_{xy}^L$, as defined in Eq.~\ref{eq:Sxymatric},
\begin{equation}
    S_{xy}^L = \frac{1}{\sum_i p_{\text{T},i}} \sum_i \frac{1}{p_{\text{T},i}} 
    \begin{bmatrix}
    p_{x,i}^2 & p_{x,i} p_{y,i} \\
    p_{x,i} p_{y,i} & p_{y,i}^2
    \end{bmatrix},
    \label{eq:Sxymatric}
\end{equation}
where the index $i$ extends over all objects used in the analysis. We require all events to have $S > 0.1$.

Another important variable is the scalar sum of the $p_T$ of the objects used in the analysis, including the missing transverse momentum. It is calculated as shown in Eq.~\ref{eq:stdef}.

\begin{equation}
    S_\text{T} = \left( \sum_{i=1}^{N} p_{\text{T},i} \right) + p_{\text{T}}^{\text{miss}}
    \label{eq:stdef}
\end{equation}

\begin{figure}[h!]
    \centering
    \includegraphics[width=0.49\linewidth]{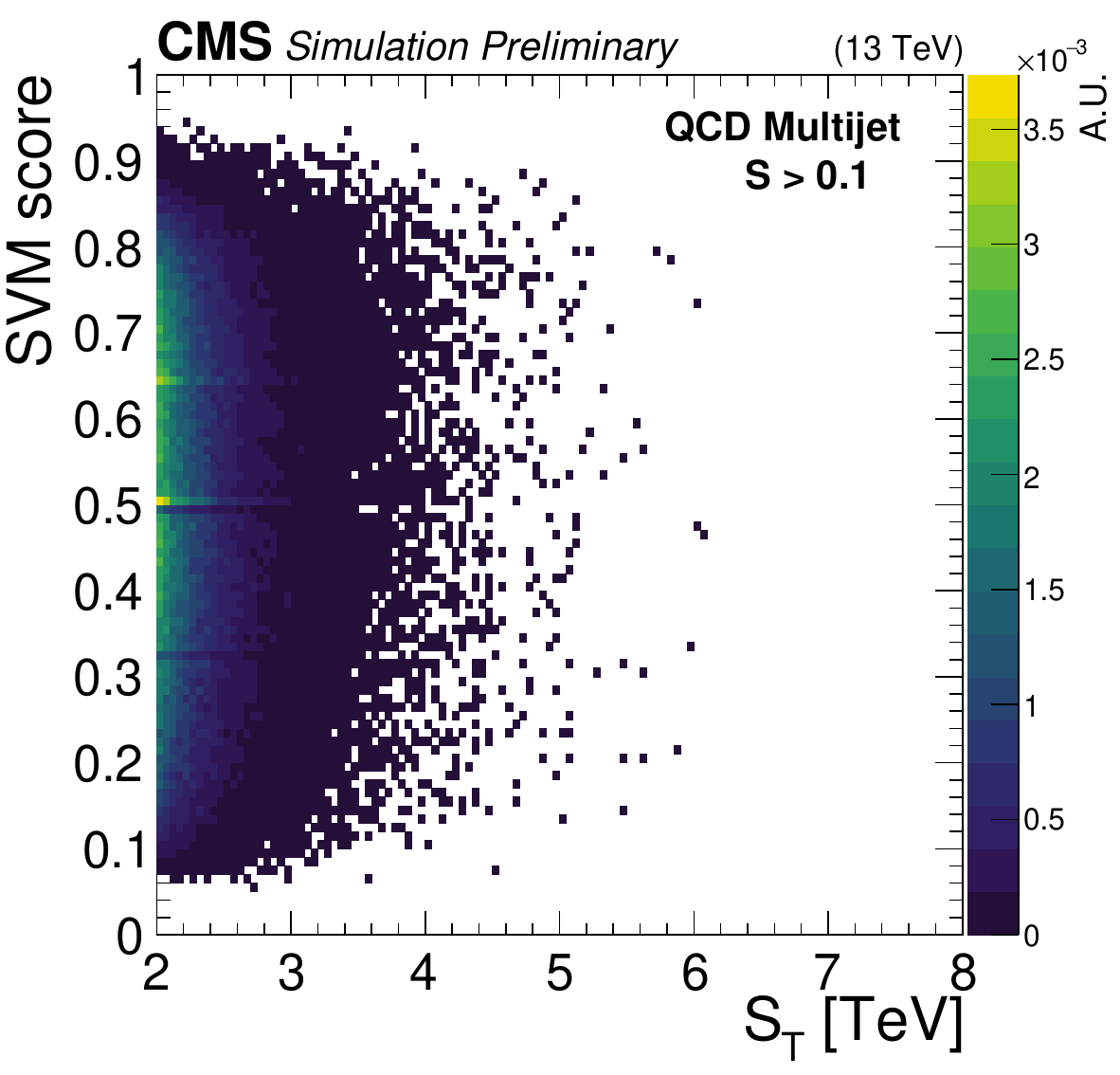}
    \includegraphics[width=0.49\linewidth]{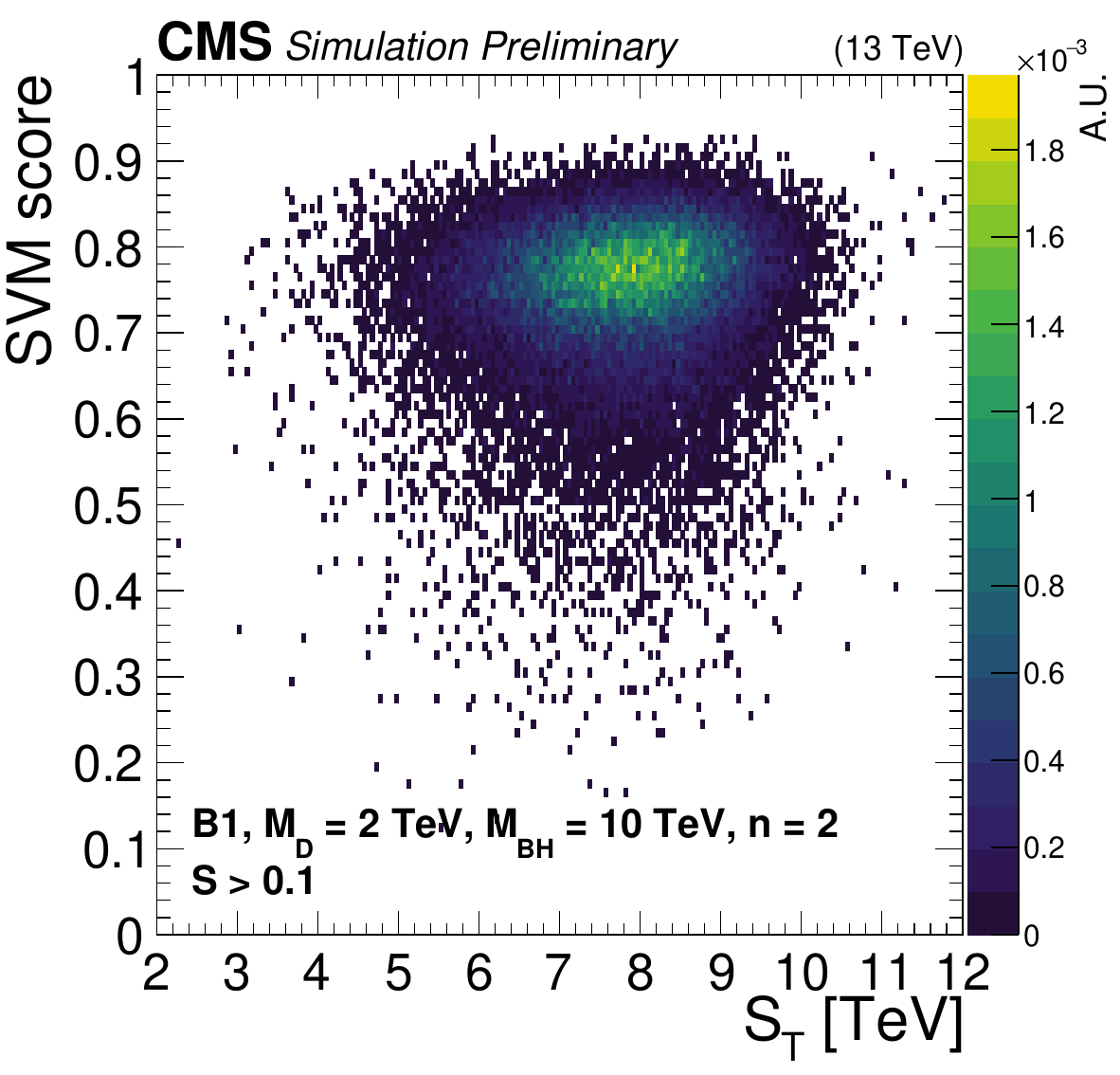}
    \caption{The SVM score vs. the $S_{T}$ distributions for simulated background (left) and a selected black hole signal model (right).}
    \label{fig:svm_vs_st}
\end{figure}

 After requiring the sphericity selection, the SVM score vs. $S_T$ distributions are shown in Fig.~\ref{fig:svm_vs_st} for simulated QCD multijet background (left) and a selected black hole signal with $M_{D} = 2~\text{TeV}$, $M_{BH} = 10~\text{TeV}$ and $n = 2$ (right). A similarly great signal to background discrimination can be observed for all models.

\section{Background prediction}

To avoid relying on the precise modeling of the high $S_T$ tails in the QCD simulations, we use a data-driven background prediction technique. We separate the data to a FAIL and PASS region, defined by the SVM score selection at 0.63, and perform a shape based analysis using the $S_T$ variable. We bin the $S_T$ distribution and fit in both regions simultaneously. We then calculate the predicted yields in the PASS region using Eq.~\ref{eq:rpf}, where we model the PASS-FAIL ratio,
$ R_{\text{P/F}}(j)$ as an exponential function, and $j$ denotes the bin index.

 \begin{equation}
	 N_{\text{PASS}}^{\text{bkg}}(j) = R_{\text{P/F}}(j) \, N_{\text{FAIL}}^{\text{bkg}}(j)
	 \label{eq:rpf}
\end{equation}

We tested this method in a validation region, where we inverted the selection on the sphericity ($S<0.1$), while keeping the other requirements the same. The prediction in this validation region shows good performance: the pulls between the data and prediction are centered at zero and have symmetric deviations.

\begin{figure*}[h!]
    \centering
     \includegraphics[width=0.8\textwidth]{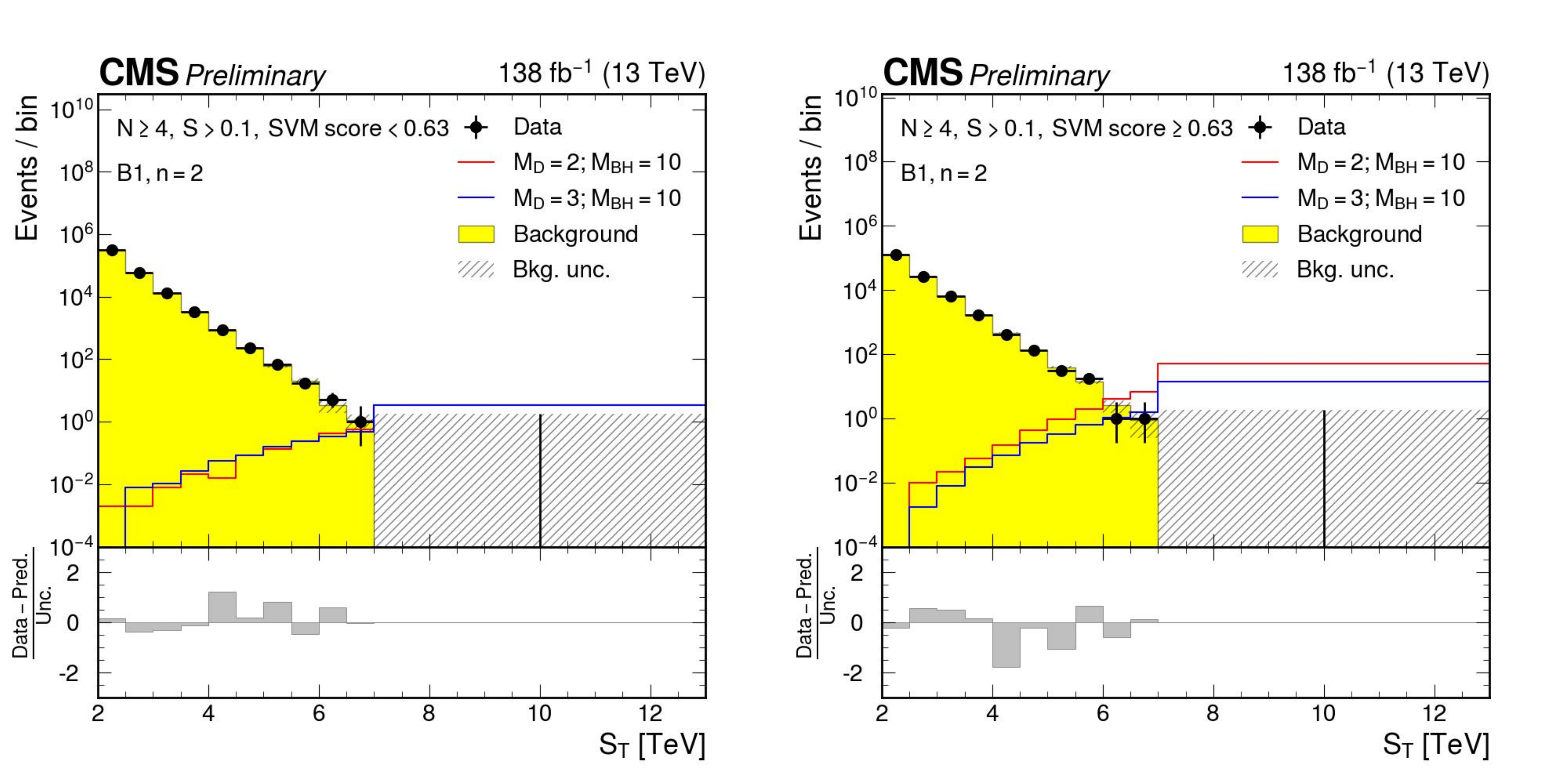}
    \caption{Post-fit $S_{T}$ distributions in the FAIL (left) and PASS (right) regions. The red and blue curves represent two selected B1 signal examples as noted in the legend. The gray shaded area shows the statistical and systematic uncertainties on the background prediction.}
    \label{fig:pass_fail_SR}
\end{figure*}

Figure~\ref{fig:pass_fail_SR} shows the post-fit $S_T$ distributions in the FAIL (left) and PASS (right) regions, together with two selected signal simulations with red and blue curves. The gray shaded area shows the statistical and systematic uncertainties on the background prediction. We do not observe any significant excess or deficit.  

\section{Results}

\begin{figure}[h!]
    \centering
     \includegraphics[width=0.45\textwidth]{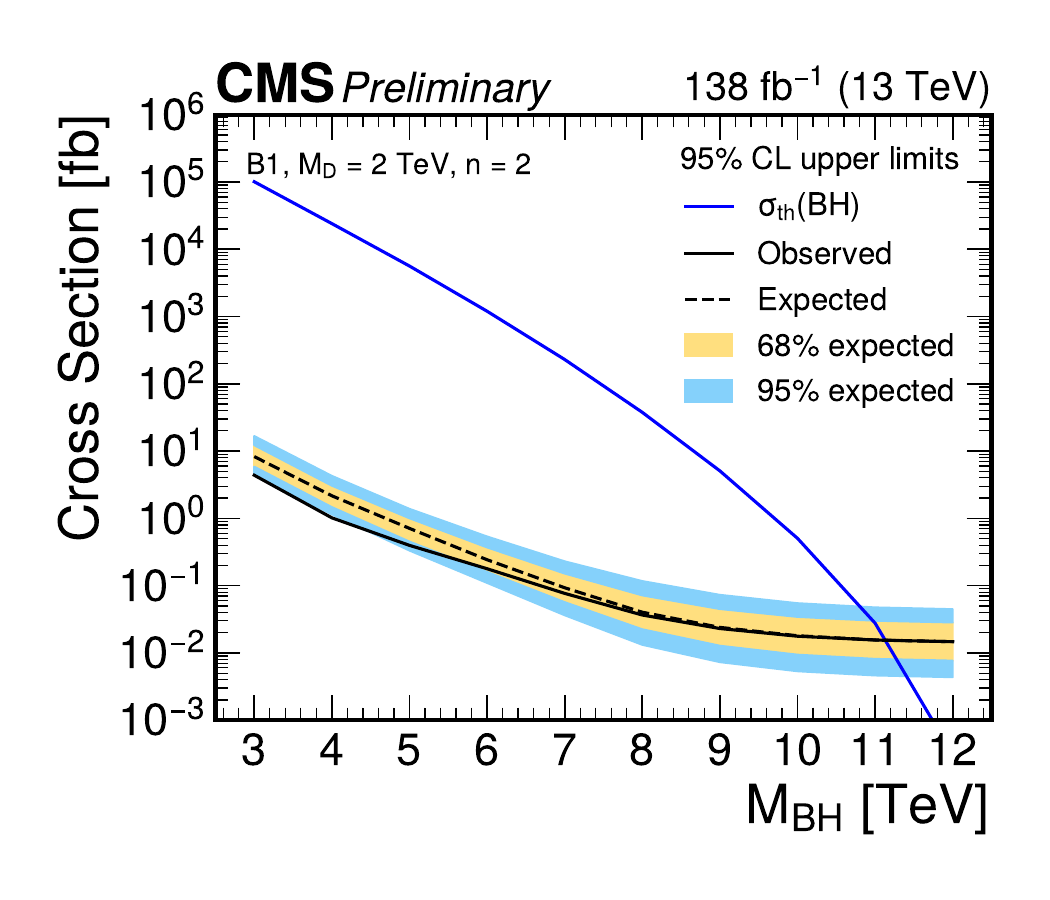}
    \caption{Expected and observed 95\% CL upper limits for B1 models with $M_{D} = 2~\text{TeV}$. The blue curves represent the theoretical cross section values.}
    \label{fig:B1_MD2_MD4}
\end{figure}

Given that we did not observe any excess or deficit, we set model-dependent 95\% CL asymptotic limits on signal cross section of the 72 BH models considered. As an example, Fig.~\ref{fig:B1_MD2_MD4} shows the expected and observed 95\% CL upper limits for B1 models with $M_{D} = 2~\text{TeV}$. The theoretical cross section is displayed with the blue curve, which includes the latest recommended parton distributions functions, \textsc{NNPDF3.1}~\cite{NNPDF:2017mvq}. The expected and observed excluded $M_\text{BH}$ is at 11.4~TeV. 

\begin{figure*}[ht!]
    \centering
    \includegraphics[width=0.49\textwidth]{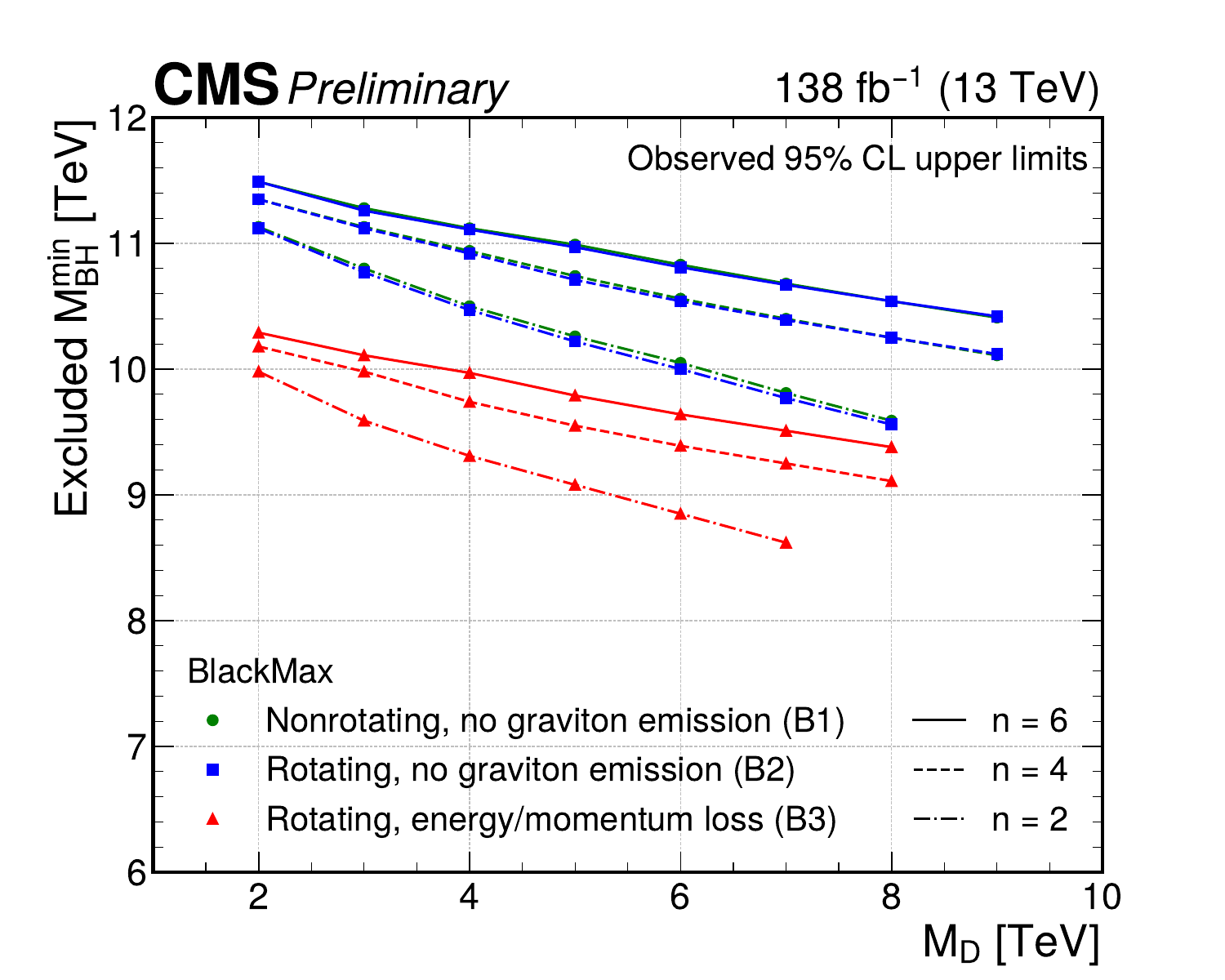}
    \includegraphics[width=0.49\textwidth]{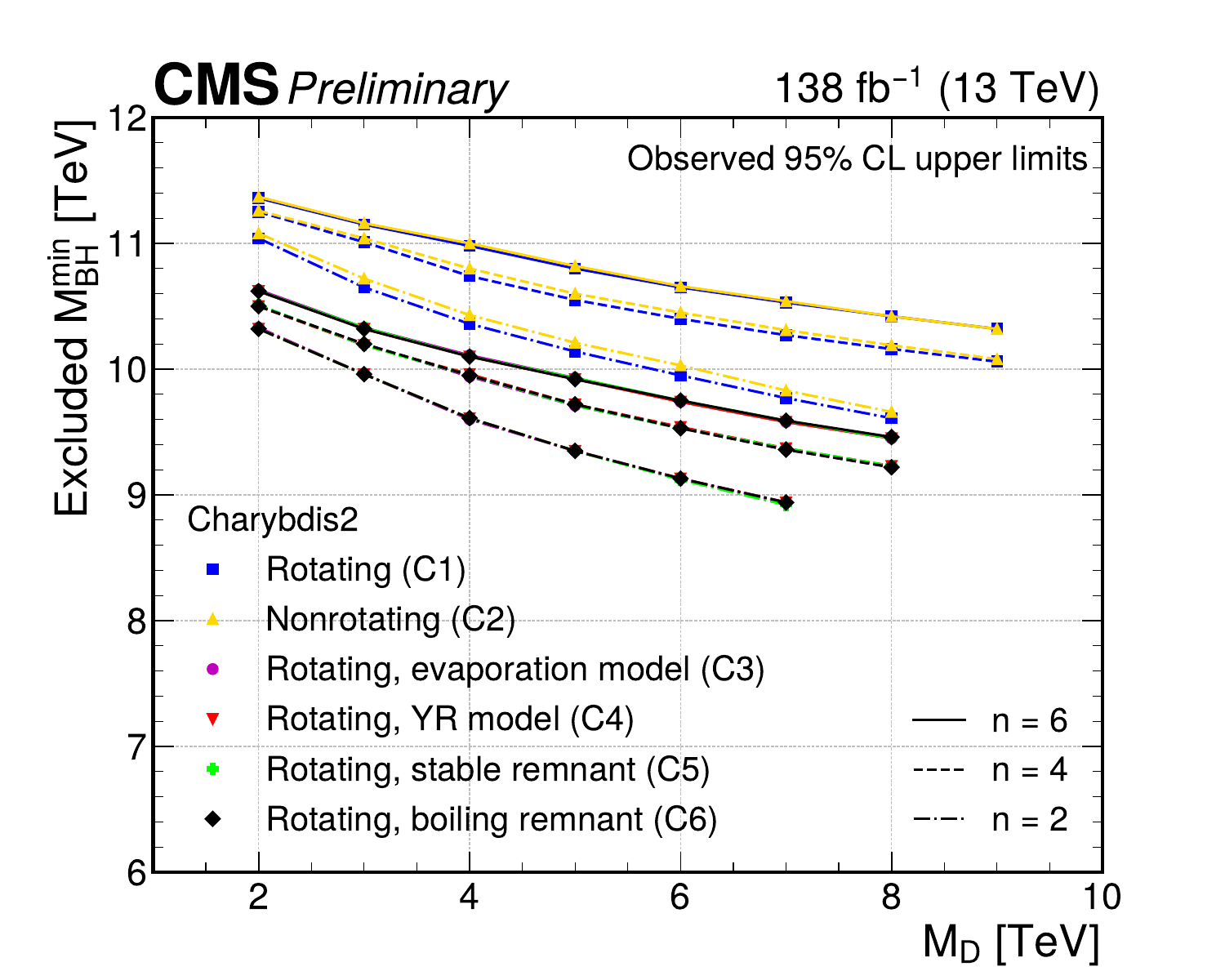}
    \caption{Excluded $M_{\text{BH}}$ values as a function of $M_{\text{D}}$ and $n$ for a variety of \textsc{BlackMax} (left) and \textsc{Charybdis2} (right) BH models.}
    \label{fig:mass_limit_blackhole}
\end{figure*}

\newpage
Following the same procedure, we can create a summary plot that shows the observed excluded $M_\text{BH}$ as a function of the $M_\text{D}$. Figure~\ref{fig:mass_limit_blackhole} shows these summary plots, in the left we see the the models with  \textsc{BlackMax}, and in the right the models with \textsc{Charybdis2}. Different extra dimension values are denoted with solid, dashed, and dash-dot styles for the curves. The colors represent the different assumptions regarding angular momentum, energy-loss and the evaporation model.

\begin{figure}[ht!]
    \centering
    \includegraphics[width=0.45\textwidth]{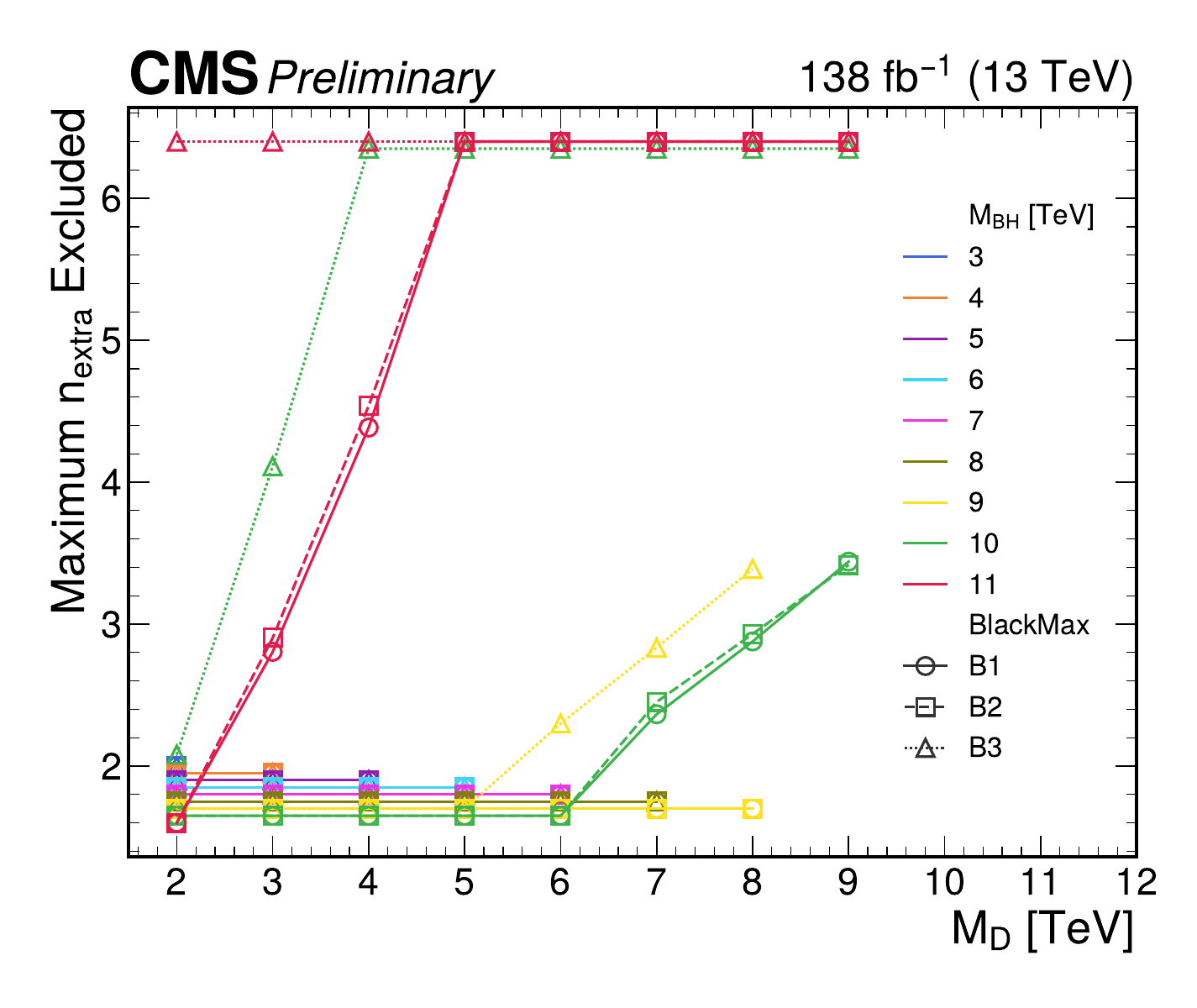}
    \caption{The maximum number of excluded extra dimensions as a function of $M_\text{D}$, $M_{\text{BH}}$ values for the \textsc{BlackMax} models. The points seen near $n_{\textrm{extra}}^{\textrm{max}}$ values of 2 and 6 are to be understood as underflow and overflow values given the simulation parameter choices at $n=2,4,6$.}
    \label{fig:extra_dim}
\end{figure}

One can look at these results as an exclusion on the allowed number of extra dimensions as well. Figure~\ref{fig:extra_dim} shows the maximum number of excluded extra dimensions as a function of $M_\text{D}$, $M_{\text{BH}}$ values for the \textsc{BlackMax} models. The results for \textsc{Charybdis2} models are very similar. These results show that for the majority of the models, the only possible extra dimension allowed is 1. The points higher than 6 are understood as an overflow, given that we did not simulate any events with higher than $n=6$.

\begin{figure}[ht!]
    \centering
    \includegraphics[width=0.49\textwidth]{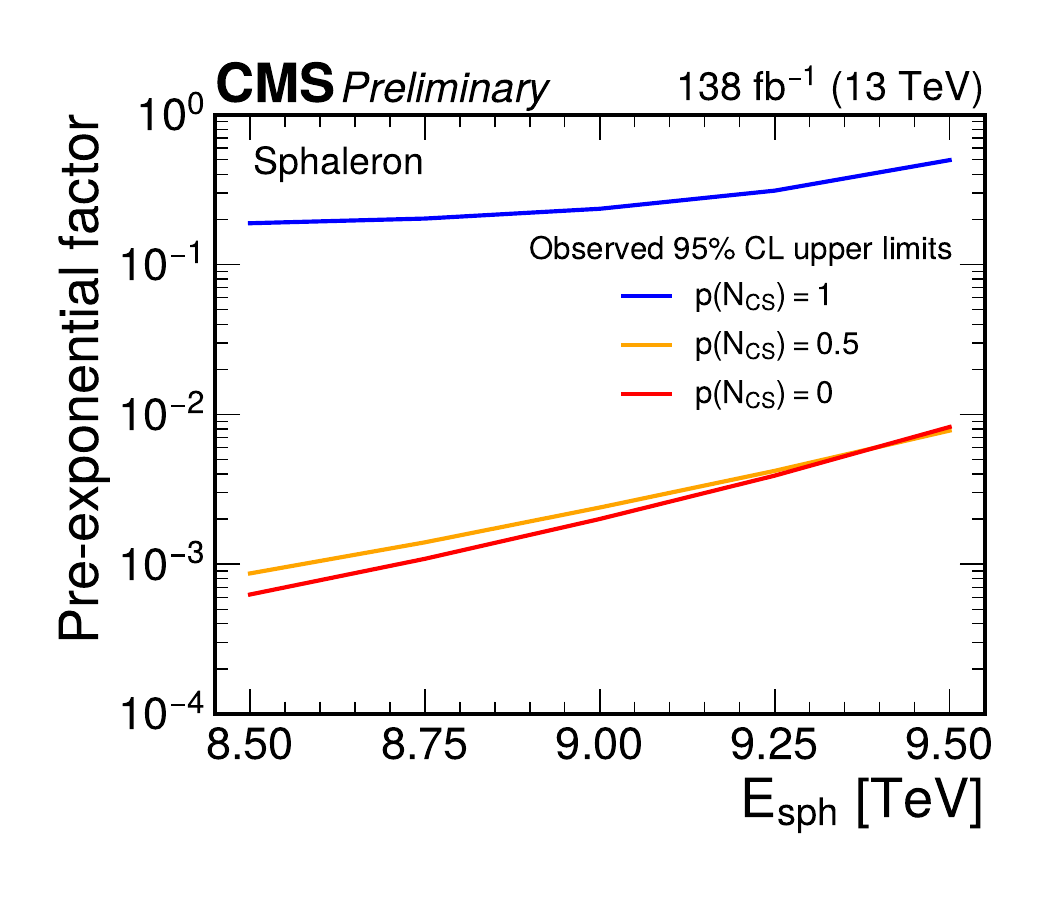}
    \caption{Observed limits on the pre-exponential factor for the sphaleron models with $p(N_{CS})=0.0$, $0.5$ and $1$.}
    \label{fig:sphaleron_limit_PEF}
\end{figure}

The last interpretation is on the sphalerons. Figure~\ref{fig:sphaleron_limit_PEF} shows the observed limits on the pre-exponential factor for the sphaleron models with $p(N_{CS})=0$, $0.5$ and $1$ as a function of the sphaleron transition energy.

\section{Summary}

We presented a new method for representing collision data and utilized advanced machine learning tools to distinguish signals from background in the search for black holes and sphalerons at the LHC, using $138~\mathrm{~fb}^{-1}$ of data collected by the CMS detector at a center-of-mass energy of 13 TeV. We excluded microscopic black holes with masses below 9.0 - 11.4~TeV, extending previous exclusions by $1{-}1.6~\text{TeV}$. We excluded the number of extra dimensions more than 1 for a variety of masses and models. We presented limits on the sphaleron pre-exponential factor to be $5\times10^{-4}$ to $2\times10^{-3}$  which is a 10-fold increase with respect to the past results, presenting the best to date results.

\bibliography{PHEP_CMS_BH}

@article{Punzi:2003bu,
    author = "Punzi, Giovanni",
    editor = "Lyons, L. and Mount, R. P. and Reitmeyer, R.",
    title = "{Sensitivity of searches for new signals and its optimization}",
    eprint = "physics/0308063",
    archivePrefix = "arXiv",
    reportNumber = "PHYSTAT-2003-MODT002",
    journal = "eConf",
    volume = "030908",
    pages = "MODT002",
    year = "2003"
}

@article{Larkoski:2020thc,
    author = "Larkoski, Andrew J. and Melia, Tom",
    title = "{Covariantizing phase space}",
    eprint = "2008.06508",
    archivePrefix = "arXiv",
    primaryClass = "hep-ph",
    doi = "10.1103/PhysRevD.102.094014",
    journal = "Phys. Rev. D",
    volume = "102",
    number = "9",
    pages = "094014",
    year = "2020"
}

@article{Cai:2024xnt,
    author = "Cai, Tianji and Cheng, Junyi and Craig, Nathaniel and Koszegi, Giacomo and Larkoski, Andrew J.",
    title = "{The phase space distance between collider events}",
    eprint = "2405.16698",
    archivePrefix = "arXiv",
    primaryClass = "hep-ph",
    doi = "10.1007/JHEP09(2024)054",
    journal = "JHEP",
    volume = "09",
    pages = "054",
    year = "2024"
}

@article{NNPDF:2017mvq,
    author = "Ball, Richard D. and others",
    collaboration = "NNPDF",
    title = "{Parton distributions from high-precision collider data}",
    eprint = "1706.00428",
    archivePrefix = "arXiv",
    primaryClass = "hep-ph",
    reportNumber = "EDINBURGH-2017-08, NIKHEF-2017-006, OUTP-17-04P, TIF-UNIMI-2017-3, CAVENDISH-HEP-17-06, CERN-TH-2017-077, Edinburgh 2017/08,
  Nikhef/2017-006, OUTP-17-04P,TIF-UNIMI-2017-3",
    doi = "10.1140/epjc/s10052-017-5199-5",
    journal = "Eur. Phys. J. C",
    volume = "77",
    number = "10",
    pages = "663",
    year = "2017"
}

@article{CMS:2008xjf,
    author = "{CMS Collaboration}",
    collaboration = "CMS",
    eprint={1003.4038},
    archivePrefix={arXiv},
    primaryClass={hep-ex},
    title = "The {CMS} experiment at the {CERN} {LHC}",
    doi = "10.1088/1748-0221/3/08/S08004",
    journal = "JINST",
    volume = "3",
    pages = "S08004",
    year = "2008"
}

@article{Ellis:2016ast,
    author = "Ellis, John and Sakurai, Kazuki",
    title = "{Search for Sphalerons in Proton-Proton Collisions}",
    eprint = "1601.03654",
    archivePrefix = "arXiv",
    primaryClass = "hep-ph",
    reportNumber = "KCL-PH-TH-2016-03, LCTS-2016-02, CERN-TH-2016-011, IPPP-16-02",
    doi = "10.1007/JHEP04(2016)086",
    journal = "JHEP",
    volume = "04",
    pages = "086",
    year = "2016"
}

@article{Randall:1999vf,
    author = "Randall, Lisa and Sundrum, Raman",
    title = "{An Alternative to compactification}",
    eprint = "hep-th/9906064",
    archivePrefix = "arXiv",
    reportNumber = "MIT-CTP-2874, PUPT-1867, BUHEP-99-13",
    doi = "10.1103/PhysRevLett.83.4690",
    journal = "Phys. Rev. Lett.",
    volume = "83",
    pages = "4690",
    year = "1999"
}

@article{Randall:1999ee,
    author = "Randall, Lisa and Sundrum, Raman",
    title = "{A Large mass hierarchy from a small extra dimension}",
    eprint = "hep-ph/9905221",
    archivePrefix = "arXiv",
    reportNumber = "MIT-CTP-2860, PUPT-1860, BUHEP-99-9",
    doi = "10.1103/PhysRevLett.83.3370",
    journal = "Phys. Rev. Lett.",
    volume = "83",
    pages = "3370",
    year = "1999"
}

@article{Arkani-Hamed:1998sfv,
    author = "Arkani-Hamed, Nima and Dimopoulos, Savas and Dvali, G. R.",
    title = "{Phenomenology, astrophysics and cosmology of theories with submillimeter dimensions and TeV scale quantum gravity}",
    eprint = "hep-ph/9807344",
    archivePrefix = "arXiv",
    reportNumber = "SLAC-PUB-7864, SU-ITP-98-142, IC-98-44",
    doi = "10.1103/PhysRevD.59.086004",
    journal = "Phys. Rev. D",
    volume = "59",
    pages = "086004",
    year = "1999"
}

@article{Antoniadis:1998ig,
    author = "Antoniadis, Ignatios and Arkani-Hamed, Nima and Dimopoulos, Savas and Dvali, G. R.",
    title = "{New dimensions at a millimeter to a Fermi and superstrings at a TeV}",
    eprint = "hep-ph/9804398",
    archivePrefix = "arXiv",
    reportNumber = "SLAC-PUB-7801, SU-ITP-98-28, CPTH-S608-0498, IC-98-39",
    doi = "10.1016/S0370-2693(98)00860-0",
    journal = "Phys. Lett. B",
    volume = "436",
    pages = "257",
    year = "1998"
}

@article{Arkani-Hamed:1998jmv,
    author = "Arkani-Hamed, Nima and Dimopoulos, Savas and Dvali, G. R.",
    title = "{The Hierarchy problem and new dimensions at a millimeter}",
    eprint = "hep-ph/9803315",
    archivePrefix = "arXiv",
    reportNumber = "SLAC-PUB-7769, SU-ITP-98-13",
    doi = "10.1016/S0370-2693(98)00466-3",
    journal = "Phys. Lett. B",
    volume = "429",
    pages = "263",
    year = "1998"
}

@article{Dai:2007ki,
    author = "Dai, De-Chang and Starkman, Glenn and Stojkovic, Dejan and Issever, Cigdem and Rizvi, Eram and Tseng, Jeff",
    title = "{BlackMax: A black-hole event generator with rotation, recoil, split branes, and brane tension}",
    eprint = "0711.3012",
    archivePrefix = "arXiv",
    primaryClass = "hep-ph",
    doi = "10.1103/PhysRevD.77.076007",
    journal = "Phys. Rev. D",
    volume = "77",
    pages = "076007",
    year = "2008"
}

@article{Harris:2003db,
    author = "Harris, C. M. and Richardson, P. and Webber, B. R.",
    title = "{CHARYBDIS: A Black hole event generator}",
    eprint = "hep-ph/0307305",
    archivePrefix = "arXiv",
    reportNumber = "CAVENDISH-HEP-03-12, CERN-TH-2003-170",
    doi = "10.1088/1126-6708/2003/08/033",
    journal = "JHEP",
    volume = "08",
    pages = "033",
    year = "2003"
}

@article{Frost:2009cf,
    author = "Frost, James A. and Gaunt, Jonathan R. and Sampaio, Marco O. P. and Casals, Marc and Dolan, Sam R. and Parker, Michael Andrew and Webber, Bryan R.",
    title = "{Phenomenology of Production and Decay of Spinning Extra-Dimensional Black Holes at Hadron Colliders}",
    eprint = "0904.0979",
    archivePrefix = "arXiv",
    primaryClass = "hep-ph",
    reportNumber = "CAVENDISH-HEP-09-05, CERN-PH-TH-2009-043",
    doi = "10.1088/1126-6708/2009/10/014",
    journal = "JHEP",
    volume = "10",
    pages = "014",
    year = "2009"
}

@article{Trodden:1998ym,
    author = "Trodden, Mark",
    title = "{Electroweak baryogenesis}",
    eprint = "hep-ph/9803479",
    archivePrefix = "arXiv",
    reportNumber = "CWRU-P6-98",
    doi = "10.1103/RevModPhys.71.1463",
    journal = "Rev. Mod. Phys.",
    volume = "71",
    pages = "1463",
    year = "1999"
}

@article{tHooft:1976rip,
    author = "'t Hooft, Gerard",
    editor = "Shifman, Mikhail A.",
    title = "{Symmetry Breaking Through Bell-Jackiw Anomalies}",
    reportNumber = "PRINT-76-0254 (HARVARD)",
    doi = "10.1103/PhysRevLett.37.8",
    journal = "Phys. Rev. Lett.",
    volume = "37",
    pages = "8",
    year = "1976"
}

@article{Klinkhamer:1984di,
    author = "Klinkhamer, Frans R. and Manton, N. S.",
    title = "{A Saddle Point Solution in the Weinberg-Salam Theory}",
    reportNumber = "NSF-ITP-84-57",
    doi = "10.1103/PhysRevD.30.2212",
    journal = "Phys. Rev. D",
    volume = "30",
    pages = "2212",
    year = "1984"
}

@article{Cortes:1995hrp,
    author = "Cortes, Corinna and Vapnik, Vladimir",
    title = "{Support-vector networks}",
    doi = "10.1007/BF00994018",
    journal = "Machine Learning",
    volume = "20",
    number = "3",
    pages = "273",
    year = "1995"
}

@article{Frederix:2018nkq,
    author = "Frederix, R. and Frixione, S. and Hirschi, V. and Pagani, D. and Shao, H. -S. and Zaro, M.",
    title = "{The automation of next-to-leading order electroweak calculations}",
    eprint = "1804.10017",
    archivePrefix = "arXiv",
    primaryClass = "hep-ph",
    reportNumber = "Nikhef/2018-015, TUM-HEP-1138/18, NIKHEF-2018-015, TUM-HEP-1138-18",
    doi = "10.1007/JHEP11(2021)085",
    journal = "JHEP",
    volume = "07",
    pages = "185",
    year = "2018",
    note = "[Erratum: JHEP 11, 085 (2021)]"
}

@article{Tye:2015tva,
    author = "Tye, S. -H. Henry and Wong, Sam S. C.",
    title = "{Bloch Wave Function for the Periodic Sphaleron Potential and Unsuppressed Baryon and Lepton Number Violating Processes}",
    eprint = "1505.03690",
    archivePrefix = "arXiv",
    primaryClass = "hep-th",
    doi = "10.1103/PhysRevD.92.045005",
    journal = "Phys. Rev. D",
    volume = "92",
    number = "4",
    pages = "045005",
    year = "2015"
}

@article{Alwall:2007fs,
    author={Alwall, J. and Hoche, S. and Krauss, F. and Lavesson, N. and Lonnblad, L. and Maltoni, F. and Mangano, M. L. and Moretti, M. and Papadopoulos, C. G. and Piccinini, F. and Schumann, S. and Treccani, M. and Winter, J. and Worek, M.},
    title = "{Comparative study of various algorithms for the merging of parton showers and matrix elements in hadronic collisions}",
    eprint = "0706.2569",
    archivePrefix = "arXiv",
    primaryClass = "hep-ph",
    reportNumber = "SLAC-PUB-12604, CERN-PH-TH-2007-066, LU-TP-07-13, KA-TP-06-2007, DCPT-07-62, IPPP-07-31",
    doi = "10.1140/epjc/s10052-007-0490-5",
    journal = "Eur. Phys. J. C",
    volume = "53",
    pages = "473",
    year = "2008"
}

@article{Papaefstathiou:2019djz,
    author = {Papaefstathiou, Andreas and Pl{\"a}tzer, Simon and Sakurai, Kazuki},
    title = "{On the phenomenology of sphaleron-induced processes at the LHC and beyond}",
    eprint = "1910.04761",
    archivePrefix = "arXiv",
    primaryClass = "hep-ph",
    reportNumber = "Nikhef 2019-045, UWTHPH-2019-30, MCnet-19-23",
    doi = "10.1007/JHEP12(2019)017",
    journal = "JHEP",
    volume = "12",
    pages = "017",
    year = "2019"
}

@article{Yoshino:2005hi,
    author = "Yoshino, Hirotaka and Rychkov, Vyacheslav S.",
    title = "{Improved analysis of black hole formation in high-energy particle collisions}",
    eprint = "hep-th/0503171",
    archivePrefix = "arXiv",
    reportNumber = "DPNU-05-01, ITFA-2005-09",
    doi = "10.1103/PhysRevD.71.104028",
    journal = "Phys. Rev. D",
    volume = "71",
    pages = "104028",
    year = "2005",
    note = "[Erratum: Phys.Rev.D 77, 089905 (2008)]"
}

@article{Koch:2005ks,
    author = "Koch, Benjamin and Bleicher, Marcus and Hossenfelder, Sabine",
    title = "{Black hole remnants at the LHC}",
    eprint = "hep-ph/0507138",
    archivePrefix = "arXiv",
    doi = "10.1088/1126-6708/2005/10/053",
    journal = "JHEP",
    volume = "10",
    pages = "053",
    year = "2005"
}

@article{Stoecker:2006we,
    author = "Stoecker, Horst",
    editor = "Bodmann, Bardo E. J. and Vasconcellos, Cesar A. Z. and Coelho, Helio T. and Hadjimichef, Dimiter and Greiner, Walter and Stucker, Horst",
    title = "{Stable TeV - Black Hole Remnants at the LHC: Discovery through Di-Jet Suppression, Mono-Jet Emission and a Supersonic Boom in the Quark-Gluon Plasma}",
    eprint = "hep-ph/0605062",
    archivePrefix = "arXiv",
    doi = "10.1142/S0218271807009930",
    journal = "Int. J. Mod. Phys. D",
    volume = "16",
    pages = "185",
    year = "2007"
}

@article{Scardigli:2008jn,
    author = "Scardigli, Fabio",
    title = "{Glimpses on the micro black hole Planck phase}",
    eprint = "0809.1832",
    archivePrefix = "arXiv",
    primaryClass = "hep-th",
    doi = "10.3390/sym12091519",
    journal = "Symmetry",
    volume = "12",
    number = "9",
    pages = "1519",
    year = "2020"
}

@article{Dimopoulos:2001qe,
    author = "Dimopoulos, Savas and Emparan, Roberto",
    title = "{String balls at the LHC and beyond}",
    eprint = "hep-ph/0108060",
    archivePrefix = "arXiv",
    reportNumber = "SU-ITP-01-36",
    doi = "10.1016/S0370-2693(01)01525-8",
    journal = "Phys. Lett. B",
    volume = "526",
    pages = "393",
    year = "2002"
}

@article{Gingrich:2008di,
    author = "Gingrich, Douglas M. and Martell, Kevin",
    title = "{Study of highly-excited string states at the Large Hadron Collider}",
    eprint = "0808.2512",
    archivePrefix = "arXiv",
    primaryClass = "hep-ph",
    doi = "10.1103/PhysRevD.78.115009",
    journal = "Phys. Rev. D",
    volume = "78",
    pages = "115009",
    year = "2008"
}

@article{Creek:2007sy,
    author = "Creek, S. and Efthimiou, O. and Kanti, P. and Tamvakis, K.",
    title = "{Greybody factors for brane scalar fields in a rotating black-hole background}",
    eprint = "hep-th/0701288",
    archivePrefix = "arXiv",
    doi = "10.1103/PhysRevD.75.084043",
    journal = "Phys. Rev. D",
    volume = "75",
    pages = "084043",
    year = "2007"
}

@article{Creek:2007pw,
    author = "Creek, S. and Efthimiou, O. and Kanti, P. and Tamvakis, K.",
    title = "{Scalar Emission in the Bulk in a Rotating Black Hole Background}",
    eprint = "0709.0241",
    archivePrefix = "arXiv",
    primaryClass = "hep-th",
    doi = "10.1016/j.physletb.2007.09.050",
    journal = "Phys. Lett. B",
    volume = "656",
    pages = "102",
    year = "2007"
}

@article{Schwarzschild:1916uq,
    author = "Schwarzschild, Karl",
    title = "{On the gravitational field of a mass point according to Einstein's theory}",
    eprint = "physics/9905030",
    archivePrefix = "arXiv",
    journal = "Sitzungsber. Preuss. Akad. Wiss. Berlin (Math. Phys. )",
    volume = "1916",
    pages = "189--196",
    year = "1916"
}

@misc{CMS-PAS-EXO-24-028,
      author        = "{CMS Collaboration}",
      collaboration = "CMS",
      title         = "{Search for microscopic black holes and sphalerons in
                       proton-proton collisions at 13 TeV}",
      institution   = "CERN",
      reportNumber  = "CMS-PAS-EXO-24-028",
      note          = "Physics Analysis Summary, CMS-PAS-EXO-24-028",
      address       = "Geneva",
      year          = "2025",
      url           = "https://cds.cern.ch/record/2945278",
}

\end{document}